# Extra-cavity-enhanced difference-frequency generation at 1.63 μm

CHEN YANG,[1, 2] SHI-LONG LIU,[1,2] ZHI-YUAN ZHOU,[1, 2, 3] * YAN LI,[1, 2] YIN-HAI LI,[3,1,2] SHI-KAI LIU,[1, 2] ZHAO-HUAI XU,[1, 2] GUANG-CAN GUO, [1,2] AND BAO-SEN SHI [1,2,3,4]

[1] CAS Key Laboratory of Quantum Information, USTC, Hefei, Anhui 230026, China
[2] Synergetic Innovation Center of Quantum Information & Quantum Physics, University of Science and Technology of China, Hefei, Anhui 230026, China
[3] Wang Da-Heng Collaborative Innovation Center for Science of Quantum Manipulation & Control, Heilongjiang Province & Harbin University of Science and Technology, Harbin 150080, China
[4] drshi@ustc.edu.cn
* zyzhouphy@ustc.edu.cn

**Abstract:** A 1632-nm laser has highly important applications in interfacing the wavelength of rubidium-based quantum memories (795 nm) and the telecom band (typically 1550 nm) by frequency conversion in three-wave mixing processes. A 1632-nm laser source based on pump-enhanced difference frequency generation is demonstrated. It has 300 mW of output power, in agreement with simulations, and a 55% quantum efficiency. An average power fluctuation of 0.51% over one hour was observed, and 200-kHz linewidth was measured using a delayed self-heterodyne method.

A quantum internet with quantum computation, communication, and metrology could extend the capabilities of telecommunication networks [1]. It would have quantum nodes connected through quantum channels. Quantum nodes are various materials systems where quantum information is generated, processed, and stored. Photons are important information carriers for transferring quantum states between remote quantum nodes. However, long-distance optical communication requires photons with wavelengths that are in the low-loss C-band, while quantum nodes usually operate at different wavelengths. Therefore, a quantum interface is needed to bridge the wavelength gap [2].

Frequency conversion was introduced to transfer qubit states from one frequency mode to another while the quantum properties are preserved [2, 3]. The most promising and relatively efficient approach is three-wave mixing in periodically poled non-linear crystals using quasi-phase-matching [4-13]. In particular, a cavity to enhance the pump field, [5, 6] or a periodically poled crystal waveguide [7, 8, 13], offers conversion efficiencies close to unity with a pump power of a few hundred milliwatts. The present work demonstrates a 1632-nm laser source that can be used as a pump to convert photons between telecom wavelengths (typically 1550 nm) and the near infrared wavelengths (795 nm) of rubidium-based quantum memories [14, 15]. The frequency up-conversion is based on sum-frequency generation (SFG) [5-9], and the corresponding down-conversion is based on difference-frequency generation (DFG) [10-13]. In addition, the long-wavelength pumping scheme can minimize background noise due to Raman scattering [8, 11].

A common method to generate lasers around 1632 nm is via semiconductor laser diodes. However, unamplified commercial 1632-nm laser diodes have low output powers around 50 mW. To obtain high power 1632nm laser, specially designed amplifiers must be used that are not readily available. Diode lasers also suffer from disadvantages such as multi-longitudinal modes and poor beam quality. In the Ref. [12], frequency down-conversion of 780-nm photons to 1522-nm photons was reported using a 1.6-μm extra-cavity diode laser as a pump. However, the quantum interface between 795 nm and telecom band has not been realized for lack of a reported pump laser. Frequency conversion based on nonlinear effects is a fundamental technique used to extend the frequency range of existing lasers [16-20]. To generate a long-wavelength laser, an optical cavity is typically used to enhance either the pump or signal fields, which leads to the implementation of pump-enhanced DFG [18, 19] and optical parametric oscillators (OPOs) [20-26], respectively. Recently, the tunable OPO with watt-level output and spanning 1.63-μm wavelengths have been reported [20-22], but no DFG schemes have been reported to date. In the DFG scheme, the signal beam is provided by external injection (instead of spontaneous generation in an OPO cavity), and thus the frequency of the generated beam is determined only by two external lasers that strictly satisfy the relation $\omega_i = \omega_p - \omega_s$ for energy conservation. Therefore, it is easier to obtain a single-mode DFG laser with low frequency noise using two high-quality signal-mode commercial laser sources. DFG allows modulation or rapid scan frequencies by tuning the signal frequency; while for an OPO cavity, the crystal temperature is changed at low speed to tune the frequency.

Here, an extra-cavity pump-enhanced DFG scheme was used to generate a 1632-nm laser by using a MgO-doped periodically poled lithium niobate (MgO: PPLN) crystal. The crystal was type-0 quasi-phase-matched with a length of 30 mm, a thickness of 1 mm, and a poled period of 20.6 μm. The two incident waves were the pump and signal waves, while the idler wave was generated at 1632 nm. The respective wavelengths were $\lambda_p$=796 nm, $\lambda_s$=1555 nm, and $\lambda_i$=1632 nm, which satisfied the relation $1/\lambda_p = 1/\lambda_s + 1/\lambda_i$. The pump laser beam was an amplified, tunable single-mode laser system, including a master oscillator (external-cavity diode laser) and a tapered amplifier. The signal laser was a continuously tunable diode laser (CTL) with a 1520–1630-nm tuning range that was amplified via an erbium-doped fiber amplifier.

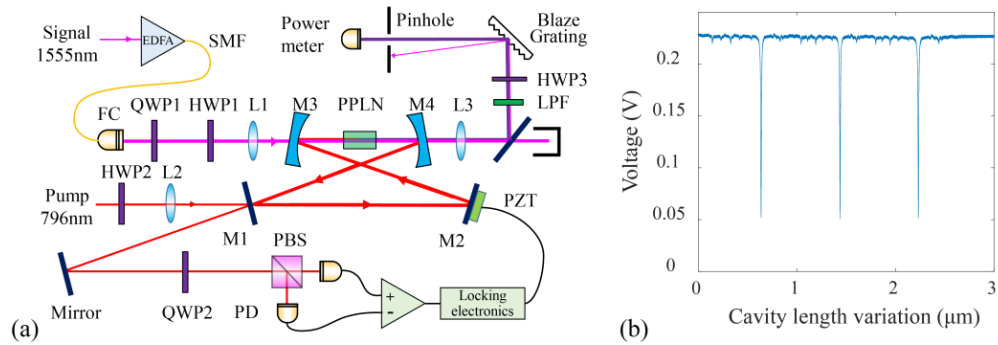

Fig. 1 (a) Schematic of the optical setup. QWP1, HWP1, and HWP2 were used to control the polarization direction to satisfy phase-matching and H3 was used to match the sensitive direction of the blazed grating. QWP2 was used in the HC locking system to generate the error signal [24]. HWP: half-wave plate; QWP: quarter-wave plate; M1-M4: cavity mirrors; L1-L3: lens with focal length of 100 mm, 500 mm, 200mm; PBS: polarizing beam splitter; SMF: single mode fiber; FC: fiber collimator; LPF: long-pass filter; DM: dichroic mirror; PZT: piezoelectric transducer; PD: photon detector. (b) Reflected spectrum of the cavity with the crystal recorded using an oscilloscope by scanning the cavity.

Fig. 1 (a) is a schematic of the setup, which can be divided into three parts: a bow-tie cavity, a homemade Hänsch-Couillaud (HC) locking system [27], and a spectral filtering stage. The cavity was comprised of four mirrors and locked to the frequency of the pump laser while the signal laser was single-passed. The input coupling mirror M1 had a power reflectivity of $R = 97.2\%$, and mirrors M2, M3, and M4 had high reflectivities (>99.9%) around 796 nm. Cavity mirrors M3 and M4 were concave with a 100-mm radius of curvature and high transmittance (>99%) around 1.63 μm and 1.55 μm, respectively. The mirror M2 was attached to a piezoelectric transducer for passive cavity stabilization based on HC-locking. The total length of the cavity was 540 mm, and the distance between M3 and M4 was 130 mm. The calculated pump beam waist at the center of the crystal was 45 μm. The signal beam was focused with a 100-mm focal length lens with a 52-μm beam waist. In the spectral filtering stage, the long-pass filter removed the pump beam and the dichroic mirror reflected the 1632 nm beam generated in the cavity while transmitting the 1555 nm beam, which was discarded. The blazed grating and the pinhole filtered the 1632 nm beam further before it was measured by a power meter.

The reflected spectrum of the cavity with the crystal, shown in Fig. 1 (b), was recorded with an oscilloscope as the cavity length was scanned. The input pump power injected was very low to reducing thermal effects when recording the reflected spectrum. From the spectrum, it was determined that the fineness $F$ was approximately 126; thus, the total loss was $\gamma \approx 5.0\%$, according to the relation $\gamma \approx 2\pi/F$. Because the input coupling mirror M1 had a power transmittance of 2.8%, the remaining loss was approximately 2.2%. In this case, the calculated impedance matching coefficient of the pump beam was 99% in theory [28]. When the cavity was locked, the transmittance of M1 was the total matching coefficient, which was the product of the impedance-matching and mode-matching coefficients. We therefore determined that the mode-matching coefficient of the pump beam was 80%, according to the reflected spectrum.

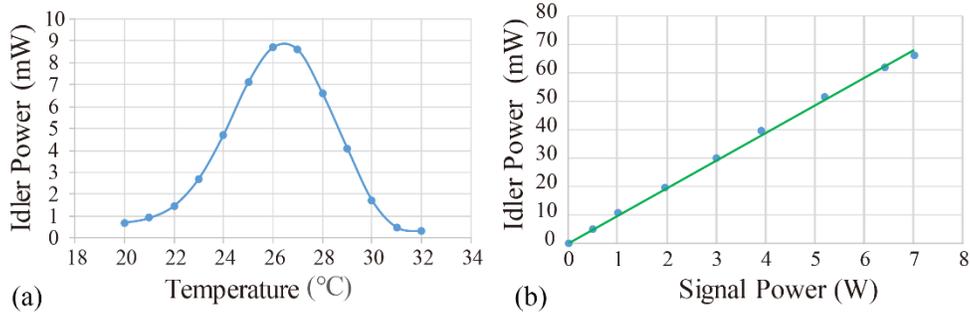

Fig. 2. Single-pass experimental data. (a) Temperature tuning curve for the single-pass configuration. (b) Single-pass DFG power as a function of the input signal power. Blue dots denote the experimental data, and the green curve was fitted with a proportional function.

Before implementing the pump-enhanced DFG, the phase-matching temperature of the crystal and the single-pass DFG efficiency were determined. Single-pass DFG was achieved by simply removing M1. When the input mirror was removed, the input pump beam was reflected by M2 and M3 and directly entered the crystal. The crystal was fixed in an oven with a temperature controlled with a semiconductor Peltier device. The DFG power as a function of crystal temperature is shown in Fig. 2 (a). The phase-matching temperature and the temperature bandwidth were 26.5 ℃ and 5 ℃, respectively. After setting the crystal temperature to 26.5 °C, the single-pass DFG power was measured as a function of the input signal power at a fixed pump power of 1.45 W. The results are shown in Fig. 2 (b), which indicates a highly linear relation that was fit with a proportional function (green line). Under the focusing conditions, the nonlinear power conversion coefficient defined as $\eta = P_i / P_p P_s \times 100\%$ was 0.67% $W^{-1}$ according to the fitting result. The optical elements of the spectral filtering stage after the crystal had a total loss of 27%, which implied that the power conversion coefficient in the crystal was $\eta_c$=0.92% $W^{-1}$. The loss was measured using a 1630-nm beam from the CTL laser.

The conversion efficiency was relatively low for single-pass DFG. The pump power could be enhanced significantly using a cavity, depending on the transmission of the input coupler and the internal cavity losses. The enhancement factor was determined by

$$\kappa = \frac{P_c}{P_p} = \frac{T}{\left[1-\sqrt{(1-T)(1-L)(1-\Gamma)}\right]^2}$$ [28], where $P_c$ was the circulating pump power in the cavity, $T = 1-R$ represented the M1 transmittance, $L = 2.7\%$ represented the round-trip linear loss (without $T$), and $\Gamma$ represented the nonlinear loss. Here, the nonlinear loss was mainly attributed to the conversion from pump photons to signal and idler photons in the DFG process and could be represented by $\Gamma = \eta_Q P_s$, where $\eta_Q = \eta_c \lambda_i / \lambda_p = 1.9\% \ W^{-1}$ was the quantum conversion coefficient.

The enhancement factor $\kappa(P_s, T)$ and the expected DFG power $P_i = \kappa \eta_c P_s P_p$ were simulated as a function of signal power $P_s$ for different $T$ without counting the mode-matching factor and the loss. The results shown in Figs. 3 (a,b) revealed an interesting fact that the cavity with lower fineness had a higher DFG power, while a higher signal power was needed. This could be attributed to the changing enhancement factor shown in Fig. 3 (a). Although the fineness of the cavity decreased with increasing transmittance $T$, the impedance matching was better because of the nonlinear loss when the signal power was high. Therefore, the enhancement factor increased with increasing transmittance $T$ for high signal power, in contrast to the opposite case for low signal power.

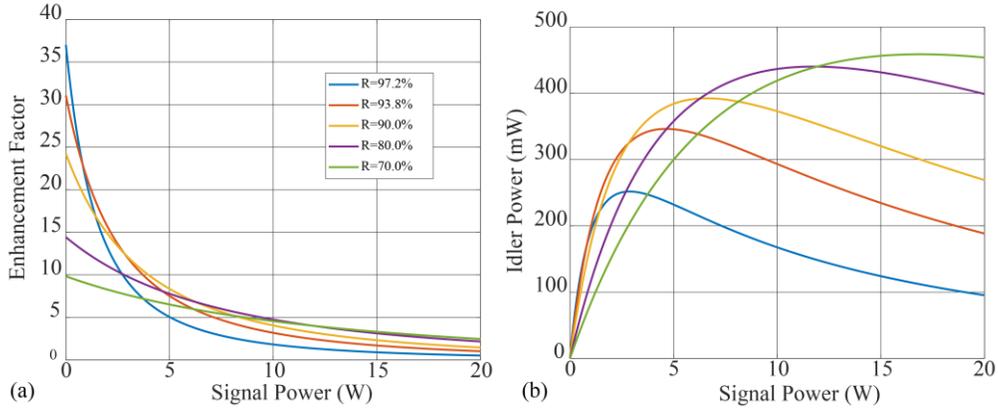

Fig. 3. Simulation results. The enhancement factor (a) and the DFG idler power (b) as a function of input signal power with fixed pump power $P_p = 1W$.

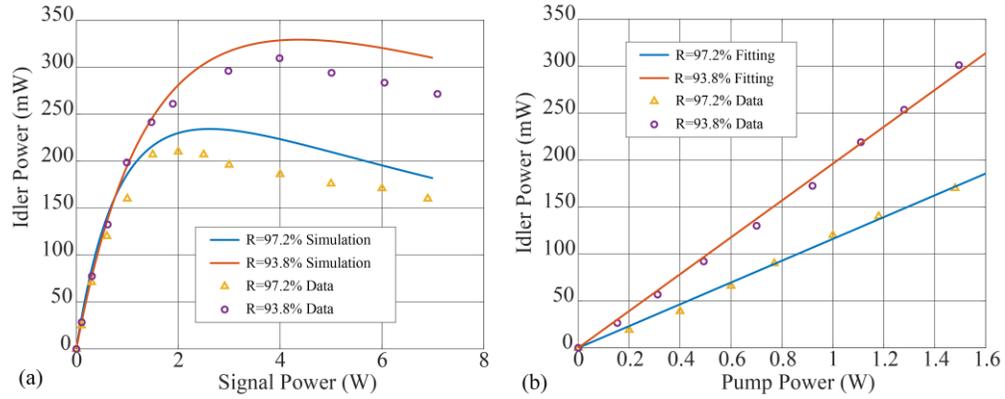

Fig. 4. Idler beam power generated from the system as function of signal power (a) and pump power (b). The blue and orange lines represented the simulation result (a) or fitting result (b) for

$R = 97.2\%$ and $R = 93.8\%$, respectively. In Fig. (a), the pump power was fixed at 1.45 W. In Fig. (b), the signal power was fixed at 1 W and 4 W in the two cases, respectively.

Two different input coupling mirrors with reflectivities of 97.2% and 93.8% were used to implement pump-enhanced DFG. The output idler power was measured as function of pump and signal powers, as shown in Figs. 4 (a,b), respectively. In Fig. 4 (a), the injected pump power was fixed at 1.45 W before M3. The solid lines represent simulation results which took account of the loss of the spectral filtering stage and the mode-matching coefficient. The data fit the simulation well in the low signal power region. The measured maximum available power was 310 mW at a signal power of 4 W and $R = 93.8\%$. This implied that a power of 425 mW was generated in the crystal. The experimental data deviated from the simulation results in the high signal power region. The differences between the two regions may be attributed to several factors, including absorption, thermal effects, and other nonlinear processes in the high-power case. In the experiment, visible green light was observed from SFG, and purple light from second harmonic generation. The differences between the single-pass and cavity-enhanced cases may also lead to differences between simulation and experimental results. To characterize the beam quality of the 1632-nm laser, the output laser beam was coupled into a single-mode fiber, with a coupling efficiency greater than 80%. This indicated good beam quality; the coupling efficiency for a diode laser would be less than 50%.

In Fig. 4 (b), the signal power was fixed at 1 W and 4 W, respectively, in the cases of $R = 97.2\%$ and $R = 93.8\%$. The linear data was fit by the proportion function. According to the fitting results, the overall power conversion efficiencies were 11.6% and 19.6%, and the quantum conversion efficiencies were 23.8% and 40.2%, respectively. Moreover, compared with the single-pass case in Fig. 2 (b), the enhancement factors were 17.3 and 7.3, respectively. Without the spectral filter stage, the internal quantum efficiency approached 55%. Using a M1 with a lower reflectivity and adjusting the pump beam for better mode-matching could improve conversion efficiency even further. In addition, according to the linear relation, the idler power could still be improved by increasing the pump power. The pump-enhanced scheme could be extended as an intra-cavity DFG laser if a gain medium for the pump beam was in the cavity; this could also improve the output power [29, 30].

The long-duration power stability and linewidth of the output beam are shown in Figs. 5 (a,b), respectively. In Fig. 5 (a), the output power over an hour was measured with power meter and sampled once per second. The relative root-mean-square power fluctuation was 0.51%. The cavity was fixed on a platform without vibration isolation, thus it was sensitive to vibrations. Two such disturbances were observed in Fig. 5 (a), yet the cavity stability was restored soon afterwards, which implies our HC locking system is reliable. In Fig. 5 (b), the spectral density was plotted by using the delayed self-heterodyne method shown schematically in Fig. 5 (c) [31]. Specifically, the output laser beam passed through an acousto-optic modulator that frequency-shifted the first-order diffracted beam by 80 MHz. The zero-order beam was coupled into a 10-km single-mode fiber for the delay and combined with the frequency-shifted diffracted beam at a 50:50 beam splitter. The two beams were then coupled into a fiber for high-speed fiber-optic detection; the resulting beat signal was displayed on a spectrum analyzer. The measured 3 dB bandwidth was 280 kHz, and the spectrum bandwidth was at least $\sqrt{2}$ times the optical linewidth [31]. Therefore, the 1632-nm laser linewidth was no more than 200 kHz.

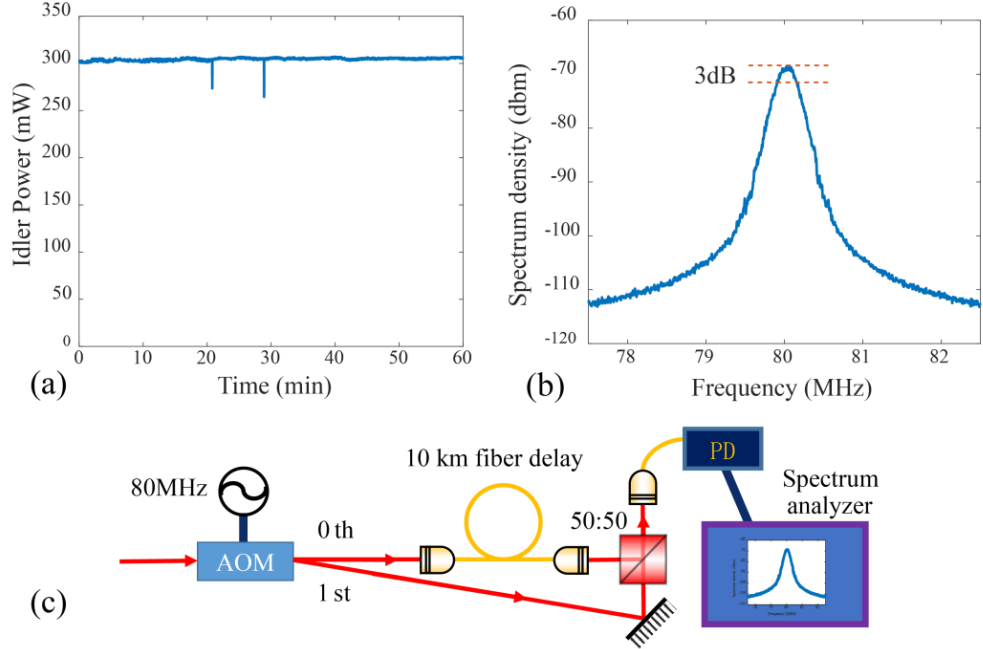

Fig. 5 (a) The output power over one hour. The root-mean-square power fluctuation was 0.51 %. (b) The optical current spectrum density from a self-heterodyne measurement. The dotted lines show the value that is 3 dB lower than the peak value. (c) Schematic of the self-heterodyne measurement.

In conclusion, a 300-mW 1632-nm laser was generated by pump-enhanced DFG. It could be used as an interface for the 795-nm D1 line of atomic rubidium and the 1550-nm telecom band. Experiments also implied that the present crystal could be adapted for a future quantum interface as long as a suitable cavity is designed. Wide wavelength tunability is not required for some atomic-based experiments, although the pump-enhanced DFG system allows the idler frequency to be tuned at will by modulating the signal laser. Thus, the system is very suitable for spectroscopy [18, 19]. Finally, the 1632-nm laser could also be used to generate mid- and far-infrared lasers by mixing with lasers in the near-infrared.

**Funding.** National Natural Science Foundation of China (NSFC) (61605194, 61435011, 61525504); Anhui Initiative In Quantum Information Technologies (AHY020200); China Postdoctoral Science Foundation (2017M622003, 2018M642517).
**Disclosures.** The authors declare no conflicts of interest.

## References

1. H J Kimble, "The quantum internet," Nature, 453(7198):1023-1030 (2008).
2. S. Tanzilli, W. Tittel, M. Halder, O. Alibart, P. Baldi, N. Gisin, and H. Zbinden, "A photonic quantum information interface," Nature 437, 116–120 (2005).
3. P. Kumar, "Quantum frequency conversion," Opt. Lett. 15, 1476–1478 (1990).
4. M.M. Fejer, G.A. Magel, D.H. Jundt, and R.L. Byer, "Quasi-phasematched second harmonic generation: tuning and tolerances," IEEE Journal of Quantum Electronics, 28(11):2631-2654, 1992.
5. Marius A. Albota and Franco N. C. Wong, "Efficient single-photon counting at 1.55 μm by means of frequency upconversion," Opt. Lett. 29, 1449-1451 (2004)
6. Aiko Samblowski, Christina E. Vollmer, Christoph Baune, Jaromír Fiurášek, and Roman Schnabel, "Weak-signal conversion from 1550 to 532 nm with 84% efficiency," Opt. Lett. 39, 2979-2981 (2014)
7. Rostislav V. Roussev, Carsten Langrock, Jonathan R. Kurz, and M. M. Fejer, "Periodically poled lithium niobate waveguide sum-frequency generator for efficient single-photon detection at communication wavelengths," Opt. Lett. 29, 1518-1520 (2004)

8. J. S. Pelc, L. Ma, C. R. Phillips, Q. Zhang, C. Langrock, O. Slattery, X. Tang, and M. M. Fejer, "Long-wavelength-pumped upconversion single-photon detector at 1550 nm: performance and noise analysis," Opt. Express 19, 21445-21456 (2011)
9. Z.-Y. Zhou, Y. Li, D.-S. Ding, W. Zhang, S. Shi, B.-S. Shi, and G.-C. Guo, "Orbital angular momentum photonic quantum interface," Light Sci Appl. 5(1), e16019 (2016).
10. Shi-Long Liu, Shi-Kai Liu, Yin-Hai Li, Shuai Shi, Zhi-Yuan Zhou, and Bao-Sen Shi, "Coherent frequency bridge between visible and telecommunications band for vortex light," Opt. Express 25, 24290-24298 (2017)
11. S. Zaske, A. Lenhard, C. A. Keßler, J. Kettler, C. Hepp, C. Arend, R. Albrecht, W.-M. Schulz, M. Jetter, and P. Michler, "Visible-to-telecom quantum frequency conversion of light from a single quantum emitter," Phys. Rev. Lett. 109(14), 147404 (2012).
12. R. Ikuta, Y. Kusaka, T. Kitano, H. Kato, T. Yamamoto, M. Koashi, and N. Imoto, "Wide-band quantum interface for visible-to-telecommunication wavelength conversion," Nature Communication. 2, 1544 (2011).
13. Sebastian Zaske, Andreas Lenhard, and Christoph Becher, "Efficient frequency downconversion at the single photon level from the red spectral range to the telecommunications C-band," Opt. Express 19, 12825-12836 (2011)
14. M. D. Eisaman, A. André, F. Massou, M. Fleischhauer, A. S. Zibrov, and M. D. Lukin, " Electromagnetically induced transparency with tunable single-photon pulses," Nature 438, 837 (2005).
15. Dong-Sheng Ding, Wei Zhang, Zhi-Yuan Zhou, Shuai Shi, Bao-Sen Shi, Guang-Can Guo, "Raman quantum memory of photonic polarized entanglement," Nat. Photon. 9, 332–338 (2015)
16. Shilong Liu, Zhenhai Han, Shikai Liu, Yinhai Li, Zhiyuan Zhou, Baosen Shi, "Efficient 525 nm laser generation in single or double resonant cavity," Opt. Communication, 410, 215 (2018).
17. Yan Li, Zhi-Yuan Zhou, Dong-Sheng Ding, and Bao-Sen Shi, "Low-power-pumped high-efficiency frequency doubling at 397.5 nm in a ring cavity," Chin. Opt. Lett. 12, 111901 (2014).
18. Kun Huang, Jiwei Gan, Jing Zeng, Qiang Hao, Kangwen Yang, Ming Yan, and Heping Zeng, "Observation of spectral mode splitting in a pump-enhanced ring cavity for mid-infrared generation," Opt. Express 27, 11766-11775 (2019)
19. Mark F. Witinski, Joshua B. Paul, and James G. Anderson, "Pump-enhanced difference-frequency generation at 3.3 μm," Appl. Opt. 48, 2600-2606 (2009)
20. Aliou Ly, Christophe Siour, and Fabien Bretenaker, "30-Hz relative linewidth watt output power 1.65 μm continuous-wave singly resonant optical parametric oscillator," Opt. Express 25, 9049-9060 (2017).
21. In-Ho Bae, Sun Do Lim, Jae-Keun Yoo, Dong-Hoon Lee, and Seung Kwan Kim, "Development of a Mid-infrared CW Optical Parametric Oscillator Based on Fan-out Grating MgO:PPLN Pumped at 1064 nm," Curr. Opt. Photon. 3, 33-39 (2019)
22. S. Chaitanya Kumar, R. Das, G.K. Samanta, M. Ebrahim-Zadeh, "Optimally-output-coupled, 17.5 W, fiber-laser-pumped continuous-wave optical parametric oscillator", Appl. Phys. B 102, 31 (2011).
23. Ritwick Das, S. Chaitanya Kumar, G. K. Samanta, and M. Ebrahim-Zadeh, "Broadband, high-power, continuous-wave, mid-infrared source using extended phase-matching bandwidth in MgO:PPLN," Opt. Lett. 34, 3836-3838 (2009)
24. Q. Sheng, X. Ding, C. Shi, S. Yin, B. Li, C. Shang, X. Yu, W. Wen, and J. Yao, "Continuous-wave mid-infrared intra-cavity singly resonant PPLN-OPO under 880 nm in-band pumping," Opt. Express 20, 8041-8046 (2012).
25. V. O. Smolski, H. Yang, S. D. Gorelov, P. G. Schunemann, and K. L. Vodopyanov, "Coherence properties of a 2.6-7.5 μm frequency comb produced as a subharmonic of a Tm-fiber laser," Opt. Lett. 41, 1388-1391 (2016).
26. Y. Li, Z. Ding, P. Liu, G. Chen, and Z. Zhang, "Widely tunable, continuous-wave, intra-cavity optical parametric oscillator based on an Yb-doped fiber laser," Opt. Lett. 43, 5391-5394 (2018).
27. T. Hansch, and B. Couillaud, "Laser frequency stabilization by polarization spectroscopy of a reflecting reference cavity," Opt. Commun. 35, 441 (1980).
28. A. Ashkin, G. D. Boyd, and J. M. Dziendzic, "Resonant optical second harmonic generation and mixing," IEEE J. Quantum Electron. 2, 109 (1966).
29. T. B. Chu and M. Broyer, "Intracavity cw dierence frequency generation by mixing three photons and using Gaussian laser beams," J. Phys. 45, 1599 (1984).
30. I. Galli, S. Bartalini, S. Borri, P. Cancio, G. Giusfredi, D. Mazzotti, and P. De Natale, "Ti:sapphire laser intracavity difference-frequency generation of 30 mW cw radiation around 4.5μm," Opt. Lett. 35, 3616-3618 (2010).
31. T. Okoshi, K. Kikuchi, and A. Nakayama, "Novel method for high resolution measurement of laser output spectrum," Electron Lett. 16, 630-631 (1980).